\documentclass[conference,a4paper]{IEEEtran}%
\usepackage[utf8]{inputenc}
\usepackage{graphicx}
\usepackage{tabularx}
\usepackage{amsmath}
\usepackage{amssymb}
\usepackage{stackengine}
\usepackage{makecell}
\usepackage{multirow}
\usepackage{cite}
\usepackage{array}
\usepackage{bbding}
\usepackage{pifont}
\usepackage{wasysym}
\usepackage{comment}
\usepackage[font=footnotesize]{caption}
\usepackage{subcaption}
\usepackage[noend]{algpseudocode}
\usepackage{color}
\usepackage{booktabs}
\usepackage{mathtools}
\usepackage{dirtytalk}
\usepackage{etoolbox}
\usepackage{tikz}
\usepackage{algorithm,algpseudocode}
\usepackage{url}
\usepackage{float}
\usepackage{stfloats}
\usepackage{eurosym}
\usepackage{cleveref}
\definecolor{headerblue}{RGB}{30, 60, 114}
\definecolor{rowgray}{RGB}{240, 242, 245}
\usepackage{titlesec}
\titlespacing*{\section}{0pt}{0.4\baselineskip}{0.2\baselineskip}


\title{Comparative Evaluation of MILP, MPC, and Reinforcement Learning for Commercial Battery Dispatch Under Time-of-Use Tariffs \vspace{-0.3cm}}

\makeatletter
\newcommand{\linebreakand}{%
  \end{@IEEEauthorhalign}
  \hfill\mbox{}\par
  \mbox{}\hfill\begin{@IEEEauthorhalign} \vspace{-12cm}
}
\makeatother
\author{
  \IEEEauthorblockN{Hafiz Majid Hussain}
  \IEEEauthorblockN{ \textit{School of Energy Systems   }}
  \IEEEauthorblockA{
       \textit{ Lappeenranta-Lahti } \\ \textit {University of Technology}\\
    Lappeenranta, Finland \\
    Majid.hussain@lut.fi
   }
  \and
  \IEEEauthorblockN{Wajiha Samar}
  \IEEEauthorblockA{ \textit {Department of Electrical}\\ \textit {and Energy Engineering}\\
   \textit{University of Vaasa,}\\
   Vaasa , Finland \\
   X9504085@student.uwasa.fi  }

 \and
 \IEEEauthorblockN{Lurian Klein}
  \IEEEauthorblockN{ \textit{Cleanwatts Digital S.A}\\
    Coimbra ,Portugal \\
   lklein@cleanwattsdigital.com
   }
\and
\IEEEauthorblockN{Pedro Nardelli}
  \IEEEauthorblockN{ \textit{School of Energy Systems   }}
  \IEEEauthorblockA{
      \textit{ Lappeenranta-Lahti } \\ \textit {University of Technology}\\
    Lappeenranta, Finland \\
   Pedro.Nardelli@lut.fi
   }
} 
\makeatletter
\patchcmd{\@maketitle}
  {\addvspace{0.5\baselineskip}\egroup}
  {\addvspace{-2\baselineskip}\egroup}
  {}
  {}
\makeatother

\begin{document}  
\maketitle
\begin{abstract}
Battery energy storage systems (BESS) paired with rooftop photovoltaics (PVs) can deliver measurable cost savings under time-of-use (TOU) electricity tariffs; however, the relative performance of model-based and model-free dispatch strategies remains insufficiently benchmarked on full-year, real-world commercial datasets. This paper presents a full-year (2023) comparative evaluation of three BESS dispatch approaches using data from a commercial PV installation operating under a TOU tariff. The examined strategies include: (i) a mixed-integer linear programming (MILP) formulation with perfect foresight, providing an oracle performance benchmark under the assumed model; (ii) a model predictive control (MPC) scheme based on a day-ahead persistence forecast, representing a low-complexity deployable approach; and (iii) a soft actor-critic (SAC) deep reinforcement learning agent trained under causal information constraints. 
The MILP benchmark achieves an annual cost reduction of 24.6\% relative to a no-storage baseline. The persistence-based MPC approach recovers 99.2\% of this benchmark using only prior-day data. In contrast, the evaluated SAC agent yields an annual cost higher than the no-storage baseline. This outcome is analyzed in the context of known challenges in reinforcement learning for energy systems, including limited observability and reward design. Overall, the results indicate that, for the studied dataset and tariff structure, persistence-based MPC captures nearly all achievable economic benefits under practical deployment constraints, whereas the considered RL configuration does not yield competitive performance under the same information limitations.
\end{abstract}

\section{Introduction}
The rapid deployment of distributed photovoltaic (PV) generation in commercial buildings, combined with the adoption of time-of-use (TOU) electricity tariffs, has increased the importance of intelligent energy management strategies. Under TOU pricing, electricity costs vary across the day, creating opportunities for battery energy storage systems (BESS) to shift energy use and reduce operational costs. In commercial PV–BESS systems, storage can be used to capture surplus solar generation and discharge during high-price periods, thereby improving the economic value of on-site generation \cite{ordonez2026photovoltaic, hussain2025hybrid}. As a result, attention has increasingly shifted toward optimizing the dispatch of already-installed storage assets.

Despite growing interest in advanced control strategies, rule-based approaches remain widely used in practice due to their simplicity, robustness, and low computational requirements \cite{majid2025crossover, erdemir2025historical, el2025efficient}. At the same time, many benchmarking studies comparing alternative dispatch methods rely on reduced time horizons, simplified datasets, or small-scale systems. In particular, relatively few studies evaluate performance over a full year using real-world commercial data, or benchmark against an oracle solution obtained under perfect foresight. Without such a reference, it is difficult to assess whether performance gaps arise from intrinsic limitations of a control method or from suboptimal implementation choices.

Existing research on BESS dispatch broadly falls into three methodological categories: optimization-based scheduling, model predictive control (MPC), and reinforcement learning (RL)-based approaches. Mixed-integer linear programming (MILP) is widely used for optimal scheduling due to its ability to explicitly represent operational constraints and economic objectives. Prior work has demonstrated the effectiveness of MILP in applications such as renewable integration and energy management under uncertainty \cite{iqbal2026risk, zheng2015optimal}. While MILP provides an upper bound on achievable performance under perfect foresight, its reliance on future information limits direct real-time applicability. To address this limitation, MPC has been extensively studied as a practical control framework. By solving a receding-horizon optimization problem using forecasted inputs, MPC can approximate long-horizon optimal solutions while remaining implementable in real time. Previous studies have shown that MPC performs well in customer-sited and renewable-integrated systems, particularly when combined with simple forecasting approaches \cite{roudnil2025energy, cortes2020practical}.

More recently, reinforcement learning (RL), especially deep RL, has been explored as a data-driven alternative for energy management under uncertainty. Algorithms such as soft actor-critic (SAC) have been applied to continuous-control energy problems, demonstrating the potential to learn control policies directly from data without requiring explicit system models \cite{zhao2026future, mohammed2026deep, haarnoja2018soft}. However, RL-based approaches often face challenges related to sample efficiency, constraint handling, and limited observability, which can impact their practical performance in real-world deployments.

Despite these advances, the literature lacks a unified benchmarking framework that simultaneously evaluates MILP, MPC, and RL-based strategies on full-year, real-world commercial datasets. Moreover, few studies compare deployable controllers against an oracle benchmark derived from perfect-foresight optimization under a consistent modeling framework. This gap limits the ability to draw robust conclusions about the relative merits of different control approaches.

This paper addresses these limitations by presenting a comprehensive full-year benchmark of three dispatch strategies applied to a real-world commercial PV–BESS system. First, a daily MILP formulation with perfect foresight is used to provide an oracle performance benchmark under the assumed model. Second, an MPC controller based on a single-day persistence forecast is implemented as a low-complexity and practically deployable solution. Third, a soft actor-critic (SAC) RL agent is trained under causal information constraints, without access to future load or generation forecasts, using a single year of historical data.

Notably, the evaluated SAC agent yields performance worse than the no-storage baseline under these conditions. Rather than treating this result as a limitation, the paper analyzes it in the context of known challenges in RL-based energy systems, including limited observability, reward design, and training data requirements. Together, the MILP benchmark, persistence-based MPC controller, and SAC evaluation provide a comparative and evidence-based assessment of controller performance for commercial PV–BESS systems under deterministic TOU tariffs.

The main contributions of this work are:
\begin{itemize}
 \item \textbf{MILP benchmark:} A full-year (35,040-interval) MILP with perfect foresight on real commercial PV–BESS data, providing an oracle reference and enabling detailed temporal performance analysis.
 \item \textbf{MPC evaluation:} A persistence-based MPC controller that achieves 99.2\% of the MILP benchmark using only prior-day data, demonstrating strong performance with minimal computational and modeling complexity.
 \item \textbf{RL assessment:} An evaluation of a SAC-based RL controller under realistic information constraints, highlighting key challenges related to observability, reward design, and data efficiency in practical BESS dispatch.
\end{itemize}
The remainder of this paper is structured as follows. Section II defines the physical system model and cost objective. Section III formulates the dispatch optimisation problem. Section IV describes each control method in detail. Section V presents experimental results. Section VI analyses the root causes of SAC RL underperformance. Section VII concludes the paper.
\section{System Model}\label{system}
This section presents the mathematical formulation of a commercial prosumer facility equipped with a grid-connected PV system and a battery energy storage system (BESS). The dispatch optimisation problem is defined over a planning horizon of $T$ discrete time steps with duration $\Delta t = 0.25\,\text{h}$ (15-minute resolution).

\subsection{Battery Energy Storage Model}

The state of charge (SoC) of the BESS at the end of time step $t$ evolves as \cite{hussain2022benchmarking,el2025efficient, hussain2018efficient}:
\begin{align}
    E_t = E_{t-1} + \eta_c \, c_t \, \Delta t - \frac{d_t}{\eta_d} \, \Delta t, \qquad t = 1, \dots, T
\end{align}
where $c_t \geq 0$ and $d_t \geq 0$ denote the charge and discharge power (kW), respectively. The SoC is bounded by the energy capacity:
\begin{align}
   0 \leq E_t \leq E^{\max}, \qquad \forall\, t.
\end{align}
The charge and discharge rates are limited by the inverter rating:
\begin{align}
0 \leq c_t \leq P^{\max}, \qquad 0 \leq d_t \leq P^{\max}, \qquad \forall\, t. \label{invert}
\end{align}
To prevent simultaneous charging and discharging, a binary variable $u_t \in \{0, 1\}$ is introduced, where $u_t = 1$ indicates charging mode:
\begin{align}
c_t \leq P^{\max} u_t, \qquad d_t \leq P^{\max} (1 - u_t), \qquad \forall\, t.
\end{align}
The initial SoC is given by $E_0 = E_{\text{init}}$ and is carried over between consecutive scheduling days.

\subsection{Energy Balance}

At each time step, the power balance at the facility bus is given by:
\begin{align}
  g_t^{\text{buy}} - g_t^{\text{sell}} = L_t - P_t^{\text{PV}} + c_t - d_t, \qquad \forall\, t \label{powerb}
\end{align}
where $L_t$ denotes the electric demand (kW), $P_t^{\text{PV}}$ the PV generation (kW), $g_t^{\text{buy}} \geq 0$ the power imported from the grid (kW), and $g_t^{\text{sell}} \geq 0$ the exported power (kW). The net load at the grid interface is:
\begin{align}
 \tilde{L}_t = L_t - P_t^{\text{PV}} + c_t - d_t.
\end{align}
A positive $\tilde{L}_t$ corresponds to grid import, while a negative value indicates export.

\subsection{Electricity Tariff Structure} \label{tou}

The facility operates under a time-of-use (TOU) retail tariff $\lambda_t$ (€/kWh) with three pricing bands:
\begin{equation}
\lambda_t =
\begin{cases}
\lambda^{\text{off}} = 0.09~\text{€/kWh}, 
& \text{off-peak } (22{:}00\!-\!08{:}00), \\

\lambda^{\text{std}} = 0.10~\text{€/kWh}, 
& \text{standard } (08{:}00\!-\!10{:}00, 13{:}00\!-\!18{:}00), \\

\lambda^{\text{pk}} = 0.20~\text{€/kWh}, 
& \text{peak } (10{:}00\!-\!13{:}00, 18{:}00\!-\!22{:}00).
\end{cases}
\end{equation}

The feed-in tariff for exported energy is fixed at $\lambda^{\text{FIT}} = 0.055\,\text{€/kWh}$, which is lower than all import price bands. This prevents profitable buy-to-export arbitrage. However, unless explicitly restricted, the optimisation model can still exploit inter-temporal arbitrage by charging the battery during low-price periods (including from the grid) and discharging during high-price periods.

\subsection{Baseline (No-Storage) Benchmark}
 
In the absence of storage, the facility operates as a net-metered prosumer. The baseline cost at each time step is:
\begin{equation}
  \phi_t^{\text{base}} = \lambda_t \max(L_t - P_t^{\text{PV}}, 0)\Delta t 
  - \lambda^{\text{FIT}} \max(P_t^{\text{PV}} - L_t, 0)\Delta t.
\end{equation}

The baseline serves as the reference for evaluating all control strategies. The total cost reduction achieved by a given method is:
\begin{align}
    \Delta C = C^{\text{base}} - C^{\text{method}},
\end{align}
and the fraction of the achievable savings captured relative to the MILP benchmark is defined as:
\begin{align}
    \eta^{\text{method}} = \frac{\Delta C}{C^{\text{base}} - C^{\text{MILP}}} \times 100\%.
\end{align}.
\vspace{-0.4cm}
\section{Problem Formulation}

The objective of the dispatch problem is to determine the optimal charging, discharging, and grid interaction decisions of the BESS so as to minimize the total electricity cost over the scheduling horizon. Given the deterministic time-of-use (TOU) tariff structure and known system constraints, the problem is formulated as a constrained optimization program over $T$ time steps. The total operational cost consists of energy imported from the grid at time-varying prices and revenue obtained from exporting excess energy under a fixed feed-in tariff. Accordingly, the optimization problem can be expressed as:
\vspace{-0.4cm}
\begin{align}
\min_{\{\mathbf{x}_t\}} \quad & \sum_{t=1}^{T} 
\left( \lambda_t \, g_t^{\text{buy}} 
- \lambda^{\text{FIT}} \, g_t^{\text{sell}} \right)\Delta t
\end{align}

subject to the battery dynamics, operational limits, and power balance constraints defined in Section~\ref{system}. The decision variables at each time step are:
\begin{align}
\mathbf{x}_t = \left\{ c_t,\; d_t,\; u_t,\; g_t^{\text{buy}},\; g_t^{\text{sell}} \right\}, \qquad \forall\, t \in \{1,\dots,T\}.
\end{align}

This formulation captures the trade-off between purchasing electricity at high-price periods and utilizing stored or locally generated energy, while accounting for battery efficiency losses and operational constraints. In the case of perfect foresight, this problem yields an oracle solution that represents the minimum achievable cost under the assumed model. In practical implementations, deviations from this benchmark arise due to limited information and forecast uncertainty.
\section{Methods}

This section describes the implementation of the considered control strategies in the context of the optimisation problem defined in Section~\ref{system}.

\subsection{MILP Upper Bound (Benchmark)}

The mixed-integer linear programming (MILP) formulation is solved using realised load and PV generation data, which are assumed known over the full horizon. This provides an oracle benchmark corresponding to the minimum achievable cost under perfect foresight, denoted by $C^{\text{MILP}}$. This benchmark is used as a reference to evaluate the performance of deployable control strategies.

To maintain computational tractability while preserving inter-day consistency, the full-year problem is decomposed into daily sub-problems of 96 time steps. The end-of-day SoC is carried forward between consecutive days:
\begin{align}
E_0^{(d+1)} = E_{96}^{(d)}.
\end{align}

The non-convex constraint preventing simultaneous charging and discharging ($c_t d_t = 0$) is enforced through an exact linearisation using a binary variable $u_t \in \{0,1\}$:
\begin{align}
 c_t \leq P^{\max} u_t, \qquad d_t \leq P^{\max}(1-u_t).
\end{align}

This results in a standard MILP formulation with 96 binary variables per day, which remains computationally modest for modern branch-and-cut solvers.

\subsection{Model Predictive Control}

Model predictive control (MPC) is implemented as a receding-horizon strategy in which an optimisation problem is solved periodically using forecasts of load and PV generation \cite{parisio2014model}. In this work, the optimisation is performed once per day over a 96-step horizon.

Since future load $L_t$ and PV generation $P_t^{\text{PV}}$ are not known in advance, a persistence forecasting model is adopted, where the next day’s profiles are assumed equal to those of the previous day:
\begin{align}
    \hat{L}_t^{(d)} = L_t^{(d-1)},  
    \hat{P}_t^{\text{PV},(d)} = P_t^{\text{PV},(d-1)}, 
    \quad t = 1,\dots,96.
\end{align}

The same optimisation formulation as the MILP benchmark is used, replacing realised values with forecasts. The resulting schedule is applied in open loop over the day without intra-day re-optimisation. Deviations from the MILP benchmark arise primarily from forecast errors and the limited information available at decision time.

Persistence-based MPC represents a low-complexity and practically deployable controller, requiring minimal data and no model training. Such approaches have been shown to achieve strong performance in similar building-scale energy management problems \cite{parisio2014model,agoundedemba2025modelling}.

\subsection{Soft Actor-Critic Reinforcement Learning} \label{SAC RL}

The BESS dispatch problem is formulated as a reinforcement learning (RL) control task in which an agent learns a policy through interaction with the environment. Under the imposed causal information structure, the agent does not have access to future load or PV generation, and therefore operates under partial observability with respect to the underlying system dynamics \cite{vazquez2019reinforcement, wicaksono2025artificial}.

\subsubsection{Environment design}
The full-year dataset is used as a sequential simulation environment with 15-minute resolution. After reaching the end of the dataset, the environment resets to the initial time step with a randomly sampled initial SoC from $[0.05E^{\max},\;0.95E^{\max}]$. This randomisation exposes the agent to a broader range of operating conditions.

\subsubsection{State space}
The state vector $\mathbf{s}_t \in \mathbb{R}^8$ includes: normalised SoC, time-of-day (encoded using sine and cosine functions), current tariff, normalised load and PV generation, and day-of-year represented using cyclic features. These encodings mitigate discontinuities in periodic time variables \cite{tang2024dynamic}.

\subsubsection{Action space and feasibility}
The action $a_t \in [-1,1]$ is mapped to charge/discharge power and clipped to enforce physical limits in real time. Constraint satisfaction is therefore handled through projection rather than explicit optimisation, and no binary decision variables are required.

\subsubsection{Algorithm}
The soft actor-critic (SAC) algorithm is employed as an off-policy actor-critic method with entropy regularisation \cite{zhang2025off}. The objective maximises expected cumulative reward together with an entropy term:
\begin{align}
\max_{\theta} \;\; \mathbb{E}_{\pi_\theta}\!\left[\sum_{t} \gamma^t \bigl(r_t + \alpha\,\mathcal{H}(\pi_\theta(\cdot|\mathbf{s}_t))\bigr)\right].
\end{align}

SAC is selected due to its sample efficiency and suitability for continuous control problems.

\subsubsection{Discount factor}
A high discount factor $\gamma = 0.999$ is used to emphasise long-term rewards, yielding $\gamma^{96} \approx 0.91$. Lower values (e.g., $\gamma = 0.99$, $\gamma^{96} \approx 0.38$) substantially reduce the relative weight of delayed rewards and can bias the policy toward short-term decisions.

\subsubsection{Training configuration}
The agent is trained for 500{,}000 environment steps using the Stable-Baselines3 library \cite{raffin2021stable}. A two-layer multilayer perceptron (256 units per layer), replay buffer of 200{,}000 transitions, batch size of 256, and a learning start threshold of 10{,}000 steps are used. The reward is defined as
\begin{align}
r_t = -\phi_t \cdot \rho,
\end{align}
with scaling factor $\rho = 10$. This formulation aligns the learning objective with cost minimisation, although it does not explicitly encode longer-term effects such as missed future arbitrage opportunities.

\subsubsection{Evaluation}
After training, the learned policy is evaluated deterministically over a full-year trajectory starting from $E_0 = E^{\max}/2$. The resulting total cost $C^{\text{RL}}$ is compared against the MPC and MILP benchmarks..

\begin{figure}
    \centering
\includegraphics[width=1\linewidth]{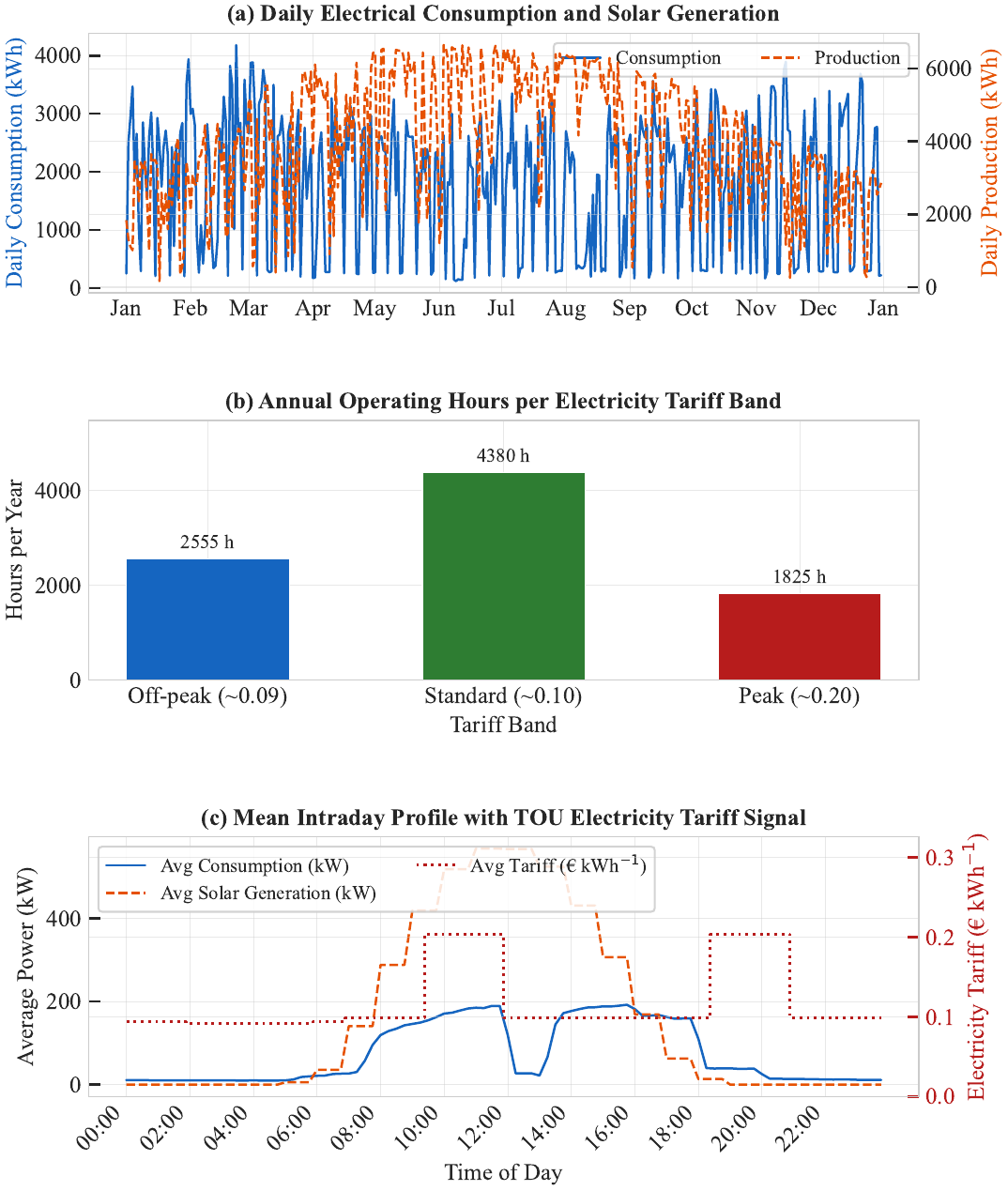}
    \caption{Exploratory analysis of the 2023 dataset: (a) daily consumption and solar generation, (b) annual hours per TOU tariff band, and (c) mean intraday power profiles overlaid with the TOU tariff signal illustrating the midday solar–peak-tariff mismatch.}
    \label{fig:baseline}
    \vspace{-0.5cm}
\end{figure}
\section{Results}

All experiments are conducted using a real-world dataset collected at a commercial facility in Portugal over the full calendar year 2023. The dataset comprises 35,040 observations at a 15-minute resolution ($\Delta t = 0.25$ h), including electricity demand $L_t$ (kW), rooftop PV generation $P_t^{\text{PV}}$ (kW), and the time-of-use (TOU) tariff $\lambda_t$ (€/kWh). The annual totals are 143.8 MWh of consumption and 68.4 MWh of PV generation, corresponding to a PV-to-load ratio of approximately 47.6\% prior to storage dispatch.

The BESS is parameterised with an energy capacity $E^{\max} = 200$ kWh, power rating $P^{\max} = 200$ kW, and charge/discharge efficiencies $\eta_c = \eta_d = 0.95$. The feed-in tariff is fixed at $\lambda^{\text{FIT}} = 0.055$ €/kWh, and the initial state of charge is $E_0 = 100$ kWh. The TOU tariff follows the three-band structure defined in Section~\ref{tou}. All optimisation problems are implemented in Python using standard optimisation libraries, and the SAC agent configuration is described in Section~\ref{SAC RL}.

\begin{figure}
    \centering
    \includegraphics[width=0.8\linewidth]{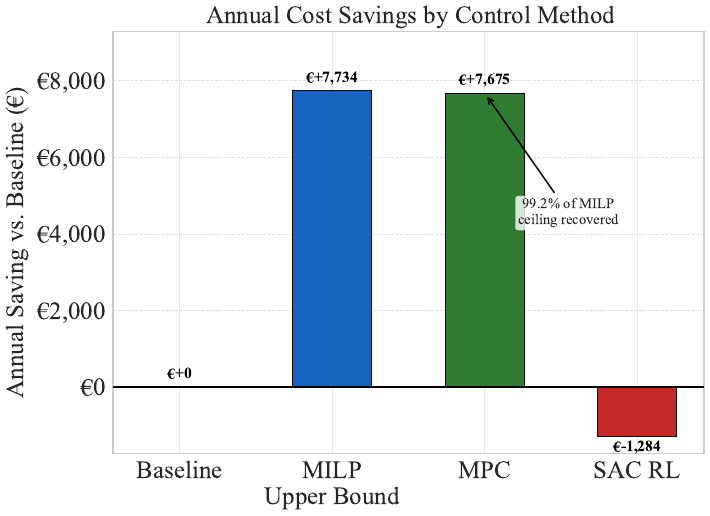}
    \caption{Annual cost savings relative to the no-storage baseline for each dispatch strategy.}
    \label{fig:annualsaving}
    \vspace{-0.8cm}
\end{figure}

Fig.~\ref{fig:baseline}(a) shows daily energy demand and PV generation over the year, highlighting strong seasonal variation, with higher solar output during summer months and increased demand during winter. Fig.~\ref{fig:baseline}(b) illustrates the distribution of hours across TOU tariff bands, where the spread between off-peak (0.09 €/kWh) and peak (0.20 €/kWh) prices creates the economic basis for energy shifting. Fig.~\ref{fig:baseline}(c) presents average intraday profiles, showing that PV generation peaks around midday, while peak tariff periods occur in the morning and evening. This temporal mismatch motivates the use of storage for cost minimisation.

Fig.~\ref{fig:annualsaving} compares annual cost savings across the three control strategies. The MILP benchmark, operating under perfect foresight, achieves an annual cost reduction of €7,733.70 (24.6\%) relative to the no-storage baseline, representing the maximum achievable savings under the assumed model. The persistence-based MPC approach attains €7,675.47, corresponding to 99.2\% of the MILP benchmark. The small performance gap (€58.23) reflects the effect of forecast inaccuracies and daily re-optimisation.

In contrast, the SAC agent results in a negative saving of €-1,283.55, yielding a higher annual cost than the no-storage baseline. This indicates that, under the considered training setup and information constraints, the learned policy does not effectively exploit TOU price differentials. The large performance gap relative to the MILP and MPC solutions suggests that limitations in state representation, reward formulation, and training data are significant factors affecting RL performance in this setting. These aspects are further analysed in the following subsection.

\begin{figure}
    \centering
    \includegraphics[width=1\linewidth]{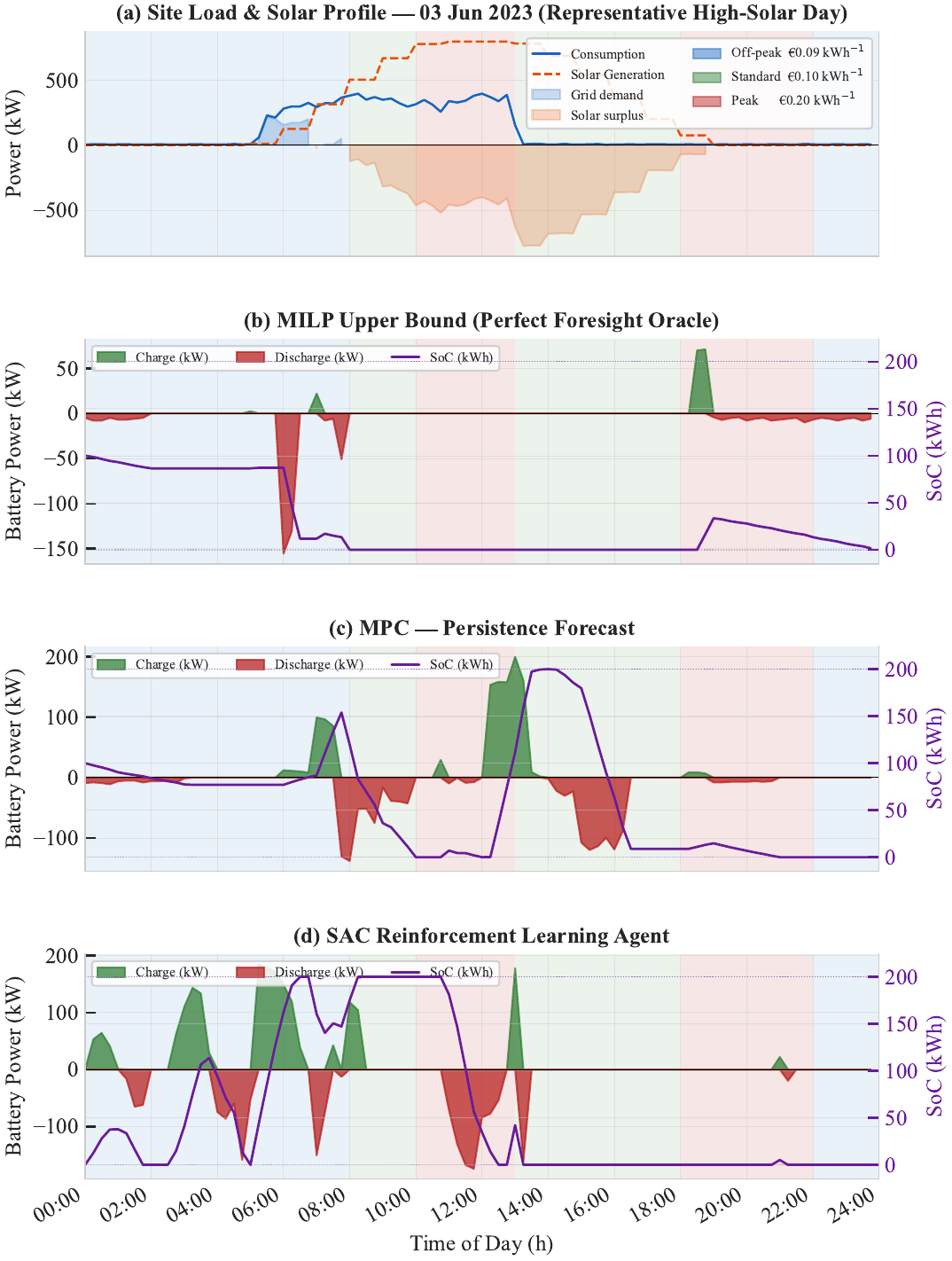}
    \caption{Intraday dispatch on a representative high-solar day (3 June 2023): (a) demand and PV generation; (b)--(d) dispatch and SoC trajectories for MILP, MPC, and SAC, respectively.}
    \label{Dispatch_da}
    \vspace{-0.8cm}
\end{figure}

Fig.~\ref{Dispatch_da} illustrates dispatch behaviour on a representative high-solar day (3 June 2023). The MILP solution follows a pre-emptive strategy, discharging the battery prior to midday to increase capacity for absorbing solar generation, and subsequently discharging during the evening peak period to avoid high-cost grid imports. The MPC controller reproduces this behaviour closely, with minor deviations in timing due to forecast error.

The SAC agent exhibits qualitatively different behaviour, including unnecessary charge--discharge cycling and suboptimal timing of energy use relative to tariff periods. These patterns indicate that the agent does not fully internalise the temporal structure of the pricing signal. As a result, the learned policy fails to capture the primary arbitrage opportunities available in the system and leads to higher overall cost.

\begin{figure}
    \centering
    \includegraphics[width=1\linewidth]{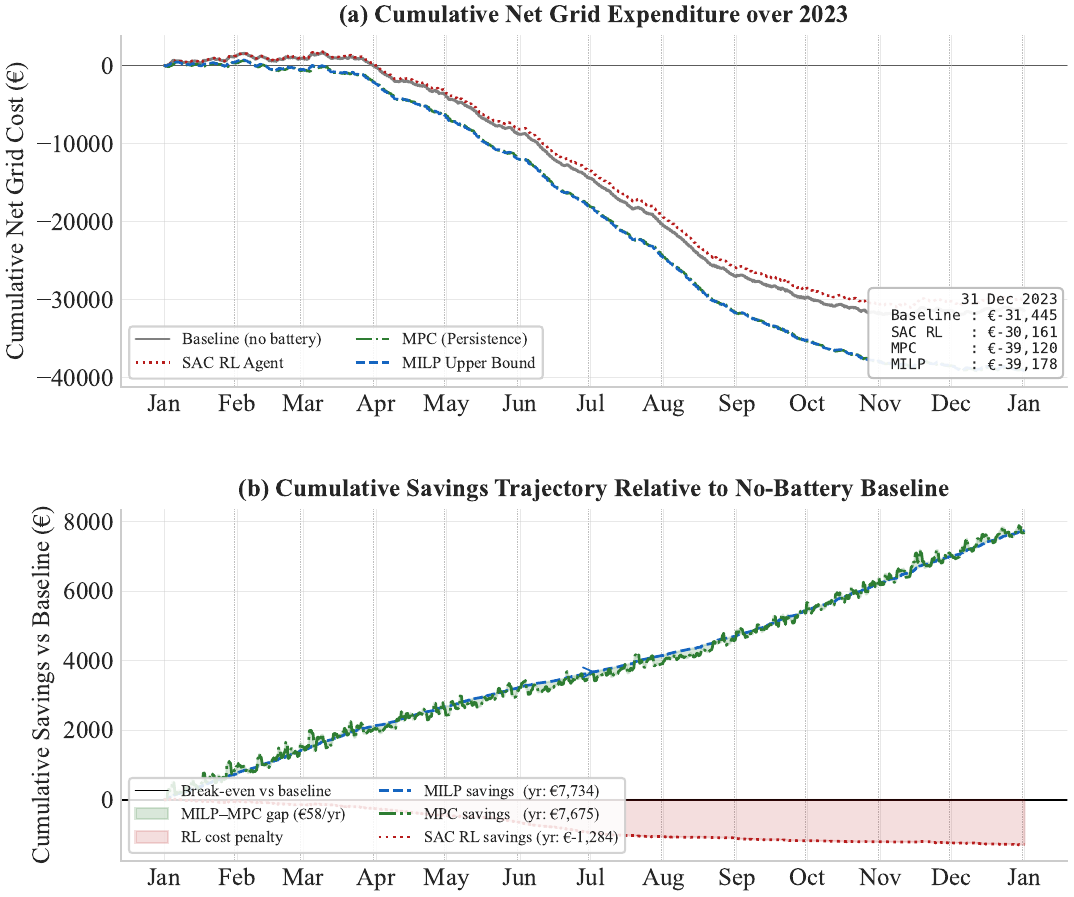}
    \caption{Full-year cumulative trajectories (Jan–Dec 2023): (a) net grid expenditure for all four strategies; (b) cumulative savings versus baseline.}
    \label{fullyear}
    \vspace{-0.5cm}
\end{figure}
Fig.~\ref{fullyear} shows the cumulative net grid expenditure and incremental savings over the year relative to the no-storage baseline. In Fig.~\ref{fullyear}(a), the MILP benchmark and MPC trajectories nearly overlap throughout the year, both achieving a substantially lower net expenditure than the baseline, indicating consistent exploitation of TOU price differences. In contrast, the SAC trajectory diverges from the baseline early in the year, resulting in higher cumulative costs. Fig.~\ref{fullyear}(b) highlights that MILP and MPC savings grow approximately linearly with only a small gap, reflecting the limited impact of forecast error under persistence. The SAC savings become negative within the first quarter and continue to decline, reaching a total penalty of €1,284 by year-end. Overall, these results show that persistence-based MPC captures nearly all achievable economic value in this setting, whereas the evaluated RL policy does not effectively exploit available arbitrage opportunities.
\subsection{Analysis of SAC RL Underperformance}

The underperformance of the SAC agent is primarily due to limitations in information availability, training setup, and reward design. First, the absence of forecast information in the state representation results in partial observability, restricting the agent’s ability to anticipate future load and PV patterns  \cite{cao2020reinforcement}. Second, the training horizon (500,000 steps) may be insufficient for stable policy learning in continuous control settings, particularly given delayed reward effects \cite{harrold2022data, cao2020reinforcement}. Third, the reward function focuses on immediate cost minimization and does not explicitly capture future opportunity costs, such as preserving stored energy for high-price periods or avoiding PV curtailment \cite{kirk2023survey}. These factors collectively limit policy quality, suggesting that improvements such as incorporating forecast features, refining reward structures, and extending training duration are necessary for achieving competitive performance.

\section{Conclusion}
This paper presented a full-year comparative evaluation of MILP, MPC, and reinforcement learning-based control strategies for battery dispatch in a commercial PV–BESS system under TOU tariffs. The MILP formulation provided an oracle benchmark under perfect foresight, enabling consistent evaluation of deployable methods. The results show that a persistence-based MPC controller achieves 99.2\% of the MILP benchmark, demonstrating that simple forecasting combined with optimisation can capture nearly all achievable economic benefits in this setting. In contrast, the evaluated SAC reinforcement learning agent yields performance below the no-storage baseline under the imposed information and training constraints. These findings highlight the effectiveness of low-complexity, model-based control for commercial solar-plus-storage systems under deterministic pricing, while also identifying key challenges for RL-based approaches, including partial observability, reward design, and training requirements. The proposed benchmarking framework provides a consistent basis for future evaluation of advanced control strategies in real-world BESS applications.

\section*{Acknowledgment}
This paper is also partly supported by (a) the Research Council of Finland through (i) ECO-NEWS n.358928, (ii) X-SDEN n.349965 and (iii) OptAd n.372723, and by (b) EU MSCA COALESCE n.101130739.
\bibliographystyle{ieeetr}
\bibliography{Ref}

@article{parisio2014model,
  title={A model predictive control approach to microgrid operation optimization},
  author={Parisio, Alessandra and Rikos, Evangelos and Glielmo, Luigi},
  journal={IEEE Transactions on Control Systems Technology},
  volume={22},
  number={5},
  pages={1813--1827},
  year={2014},
  publisher={IEEE}
}

@article{agoundedemba2025modelling,
  title={Modelling and optimization of microgrid with combined genetic algorithm and model predictive control of PV/Wind/FC/battery energy systems},
  author={Agoundedemba, Maklewa and Kim, Chang Ki and Kim, Hyun-Goo and Nyenge, Raphael and Musila, Nicholas},
  journal={Energy Reports},
  volume={13},
  pages={238--255},
  year={2025},
  publisher={Elsevier}
}

@article{vazquez2019reinforcement,
  title={Reinforcement learning for demand response: A review of algorithms and modeling techniques},
  author={V{\'a}zquez-Canteli, Jos{\'e} R and Nagy, Zolt{\'a}n},
  journal={Applied energy},
  volume={235},
  pages={1072--1089},
  year={2019},
  publisher={Elsevier}
}

@article{wicaksono2025artificial,
  title={Artificial-intelligence-enabled dynamic demand response system for maximizing the use of renewable electricity in production processes},
  author={Wicaksono, Hendro and Trat, Martin and Bashyal, Atit and Boroukhian, Tina and Felder, Mine and Ahrens, Mischa and Bender, Janek and Gro{\ss}, Sebastian and Steiner, Daniel and July, Christoph and others},
  journal={The International Journal of Advanced Manufacturing Technology},
  volume={138},
  number={1},
  pages={247--271},
  year={2025},
  publisher={Springer}
}

@article{tang2024dynamic,
  title={Dynamic demand-aware power grid intelligent pricing algorithm based on deep reinforcement learning},
  author={Tang, Chao and Qin, Yunchuan and Wu, Fan and Tang, Zhuo},
  journal={IEEE Access},
  volume={12},
  pages={75809--75817},
  year={2024},
  publisher={IEEE}
}

@article{zhang2025off,
  title={Off-policy asymptotic and adaptive maximum entropy deep reinforcement learning},
  author={Zhang, Huihui and Han, Xu},
  journal={International Journal of Machine Learning and Cybernetics},
  volume={16},
  number={4},
  pages={2417--2429},
  year={2025},
  publisher={Springer}
}

@article{raffin2021stable,
  title={Stable-baselines3: Reliable reinforcement learning implementations},
  author={Raffin, Antonin and Hill, Ashley and Gleave, Adam and Kanervisto, Anssi and Ernestus, Maximilian and Dormann, Noah},
  journal={Journal of machine learning research},
  volume={22},
  number={268},
  pages={1--8},
  year={2021}
}

@article{ordonez2026photovoltaic,
  title={Photovoltaic self-consumption in developing countries: Assessing the impact of block tariffs and net metering in Ecuador},
  author={Ord{\'o}{\~n}ez, {\'A}ngel and S{\'a}nchez, Esteban},
  journal={Energy for Sustainable Development},
  volume={92},
  pages={101955},
  year={2026},
  publisher={Elsevier}
}

@article{erdemir2025historical,
  title={Historical dimensions and directions on energy storage: unique perspectives},
  author={Erdemir, Dogan and Dincer, Ibrahim},
  journal={Journal of Energy Storage},
  volume={128},
  pages={117199},
  year={2025},
  publisher={Elsevier}
}

@article{el2025efficient,
  title={Efficient energy management of a low-voltage AC microgrid with renewable and energy storage integration using nonlinear control},
  author={El Mezdi, Karim and El Magri, Abdelmounime and Watil, Aziz and El Myasse, Ilyass and El Aadouli, Nabil and Kumar, Pankaj},
  journal={Scientific Reports},
  volume={15},
  number={1},
  pages={38651},
  year={2025},
  publisher={Nature Publishing Group UK London}
}

@article{iqbal2026risk,
  title={Risk-Aware Data Center BESS Dispatch With MILP, Machine Learning, and Stochastic Outage Simulations},
  author={Iqbal, Hasan and Sarwat, Arif},
  journal={IEEE Access},
  year={2026},
  publisher={IEEE}
}

@article{zheng2015optimal,
  title={Optimal short-term power dispatch scheduling for a wind farm with battery energy storage system},
  author={Zheng, Y and Hill, DJ and Meng, K and Luo, FJ and Dong, ZY},
  journal={IFAC-PapersOnLine},
  volume={48},
  number={30},
  pages={518--523},
  year={2015},
  publisher={Elsevier}
}

@article{cortes2020practical,
  title={Practical Considerations For Customer-sited Energy Storage Dispatch On Multiple Applications Using Model Predictive Control},
  author={Cortes, Andres and Sharma, Vinayak and Garg, Aditie and Stevens, David and Cali, Umit},
  journal={IFAC-PapersOnLine},
  volume={53},
  number={2},
  pages={12465--12470},
  year={2020},
  publisher={Elsevier}
}

@article{roudnil2025energy,
  title={Energy Management of Microgrids: An MPC-Based Techno-Economic Optimisation for RES Integration and ESS Utilisation},
  author={Roudnil, Sina and Ghassem Zadeh, Saeid and Feyzi, Mohammad Reza and Aminzadeh Ghavifekr, Amir},
  journal={IET Generation, Transmission \& Distribution},
  volume={19},
  number={1},
  pages={e70082},
  year={2025},
  publisher={Wiley Online Library}
}

@article{zhao2026future,
  title={Future Trajectory Representation-Aided DRL for Real-Time Battery Energy Storage Dispatch in Distribution Networks},
  author={Zhao, Pengfei and Hu, Weihao and Cao, Di and Du, Jialin and Zhang, Zhenyuan and Huang, Qi and Chen, Zhe},
  journal={IEEE Transactions on Smart Grid},
  year={2026},
  publisher={IEEE}
}

@article{mohammed2026deep,
  title={Deep Reinforcement Learning for Battery Energy Storage Optimization and Residential Decarbonization in Grid-Deficient Environments: An Iraqi Case Study},
  author={Mohammed, Ahmed and Abdullah, Badr M and Shubbar, Ali and Zhang, Qian and Aldhaibani, Omar and Cullen, Jeff and Salih, Amer},
  journal={Energies},
  volume={19},
  number={5},
  pages={1233},
  year={2026},
  publisher={MDPI}
}

@inproceedings{haarnoja2018soft,
  title={Soft actor-critic: Off-policy maximum entropy deep reinforcement learning with a stochastic actor},
  author={Haarnoja, Tuomas and Zhou, Aurick and Abbeel, Pieter and Levine, Sergey},
  booktitle={International conference on machine learning},
  pages={1861--1870},
  year={2018},
  organization={Pmlr}
}

@article{harrold2022data,
  title={Data-driven battery operation for energy arbitrage using rainbow deep reinforcement learning},
  author={Harrold, Daniel JB and Cao, Jun and Fan, Zhong},
  journal={Energy},
  volume={238},
  pages={121958},
  year={2022},
  publisher={Elsevier}
}

@article{cao2020reinforcement,
  title={Reinforcement learning and its applications in modern power and energy systems: A review},
  author={Cao, Di and Hu, Weihao and Zhao, Junbo and Zhang, Guozhou and Zhang, Bin and Liu, Zhou and Chen, Zhe and Blaabjerg, Frede},
  journal={Journal of modern power systems and clean energy},
  volume={8},
  number={6},
  pages={1029--1042},
  year={2020},
  publisher={SGEPRI}
}

@article{kirk2023survey,
  title={A survey of zero-shot generalisation in deep reinforcement learning},
  author={Kirk, Robert and Zhang, Amy and Grefenstette, Edward and Rockt{\"a}schel, Tim},
  journal={Journal of Artificial Intelligence Research},
  volume={76},
  pages={201--264},
  year={2023}
}

@article{hussain2025hybrid,
  title={A Hybrid Heuristic Solution for Packetized Energy Transactions in Smart Homes},
  author={Hussain, Hafiz Majid and Ahmad, Ashfaq and Nardelli, Pedro HJ},
  journal={International Journal of Energy Research},
  volume={2025},
  number={1},
  pages={3241856},
  year={2025},
  publisher={Wiley Online Library}
}

@article{majid2025crossover,
  title={Crossover-BPSO Driven Multi-Agent Technology for Managing Local Energy Systems},
  author={Majid Hussain, Hafiz and Nardelli, Ashfaq Ahmad Pedro HJ},
  journal={arXiv e-prints},
  pages={arXiv--2501},
  year={2025}
}

@article{hussain2022benchmarking,
  title={Benchmarking of heuristic algorithms for energy router-based packetized energy management in smart homes},
  author={Hussain, Hafiz Majid and Ahmad, Ashfaq and Narayanan, Arun and Nardelli, Pedro HJ and Yang, Yongheng},
  journal={IEEE Systems Journal},
  volume={17},
  number={2},
  pages={2721--2732},
  year={2022},
  publisher={IEEE}
}

@article{hussain2018efficient,
  title={An efficient demand side management system with a new optimized home energy management controller in smart grid},
  author={Hussain, Hafiz Majid and Javaid, Nadeem and Iqbal, Sohail and Hasan, Qadeer Ul and Aurangzeb, Khursheed and Alhussein, Musaed},
  journal={Energies},
  volume={11},
  number={1},
  pages={190},
  year={2018},
  publisher={MDPI}
}
\end{document}